%% file: paper.tex
\documentclass[runningheads]{llncs}

\usepackage{floatrow}
\newfloatcommand{capbtabbox}{table}[][\FBwidth]

\usepackage{blindtext}

\usepackage{graphicx}
\input{latex-includes}

\newcommand{\monosideeffect}{mono side effect\xspace}
\newcommand{\monosideeffects}{mono side effects\xspace}
\newcommand{\polysideeffect}{polypharmacy side effect\xspace}
\newcommand{\polysideeffects}{polypharmacy side effects\xspace}
\newcommand{\distmult}{\textsc{DistMult}\xspace}

\newcommand{\decagon}{\textsc{Decagon}\xspace}
\newcommand{\twosides}{\textsc{TWOSIDES}\xspace}
\newcommand{\stitch}{\textsc{STITCH}\xspace}
\newcommand{\sider}{\textsc{SIDER}\xspace}
\newcommand{\offsides}{\textsc{OFFSIDES}\xspace}

\newcommand{\kblrn}{\textsc{KBlrn}\xspace}

\begin{document}
\title{\textbf{Poster:} Knowledge Graph Completion to Predict Polypharmacy Side Effects}
\titlerunning{KG Completion to Predict Polypharmacy Side Effects}

%%%
% Author info
%%%
\author{Brandon Malone\inst{1} \orcidID{0000-0002-7027-3157} \and
Alberto  Garc{\'i}a-Dur{\'a}n\inst{1} \and %\orcidID{1111-2222-3333-4444} \and
\\ Mathias Niepert\inst{1}} %\orcidID{2222--3333-4444-5555}}
%}
%\author{Submission \#13}
%
\authorrunning{B. Malone et al.}
%\authorrunning{Submission \#13}

%%%
% Institute info
%%%
\institute{NEC Laboratories Europe, K\"urfursten-Anlage 36, 69115 Heidelberg, Germany\\
\email{\{brandon.malone,alberto.duran,mathias.niepert\}@neclab.eu}}
%\institute{}
%
\maketitle              % typeset the header of the contribution
\begin{abstract}
The \polysideeffect prediction problem considers cases in which two
drugs taken individually do not result in a particular side effect; however,
when the two drugs are taken in combination, the side effect manifests.
In this work, we demonstrate that multi-relational knowledge graph completion
achieves state-of-the-art results on the \polysideeffect prediction
problem.
Empirical results show that our approach is particularly effective
when the protein targets of the drugs are well-characterized.
In contrast to prior work, our approach provides more interpretable
predictions and hypotheses for wet lab validation.

\end{abstract}
%
%
%

%\todo{be considstent about "gene" and "protein".}

\section{Introduction}
\label{sec:introduction}

Disease and other health-related problems are often treated with medication.
In many cases, though, multiple medications may be given to treat either a
single condition or to account for co-morbidities. However, such combinations
significantly increase the risk of unintended side effects due to unknown
drug-drug interactions.

In this work, we show that multi-relational knowledge graph (KG) completion gives
state-of-the-art performance in predicting these unknown drug-drug interactions.
The KGs are multi-relational in the sense that they contain edges with different types.
We formulate the problem as a multi-relational link prediction problem in a
KG and adapt existing graph embedding strategies to
predict the interactions. 
In contrast to prior approaches for the \polysideeffect problem, we
incorporate interpretable features; thus, our approach naturally
yields explainable predictions and suggests hypotheses for wet lab validation.
Further, while we
focus on the side effect prediction problem, our approach is general and can
be applied to any multi-relational link prediction problem.

%\section{Background}
%\label{sec:background}

Much recent work has considered the problem of predicting drug-drug
interactions (e.g.~\cite{Cheng2014,Zhang2017a} and probabilistic approaches like~\cite{Sridhar2016}). However, these approaches only
consider \emph{whether} an interaction occurs; they do not consider the \emph{type
of interaction} as we do here. Thus, these methods are not directly comparable.
The recently-proposed \decagon approach~\cite{Zitnik2018} is most similar to ours;
they also predict types of drug-drug interactions. However, they use a complicated
combination of a graph convolutional network and a tensor factorization.
In contrast, we use a neural KG embedding method in combination with a method to incorporate rule-based features.
Hence, our method explicitly captures meaningful \emph{relational features}. Empirically, we demonstrate that
our method outperforms \decagon in Section~\ref{sec:results}.

\section{Datasets}
\label{sec:datasets}

\vspace{-3mm}

\begin{figure}[t!]
\begin{floatrow}
\capbtabbox{%
\scriptsize
\begin{tabular}{l|r}
                                  & \multicolumn{1}{c}{\textbf{Count}}                \\ \hline
Proteins                          &   $19~089$                   \\
Drugs                             &   $645$                   \\
Protein-protein interactions      &   $715~612$                   \\
Drug-drug interactions            &    $4~649~441$                  \\
Drug-protein target relationships &   $11~501$                   \\
Mono side effects                 &  $174~977$ \\
Distinct \monosideeffects         &  $10~184$\\
Distinct \polysideeffects         &  $963$\\
\end{tabular}
}{%
  \caption{Size statistics of the graph\label{tab:graph-size}}%
}
\ffigbox{%
  \includegraphics[width=0.86\linewidth]{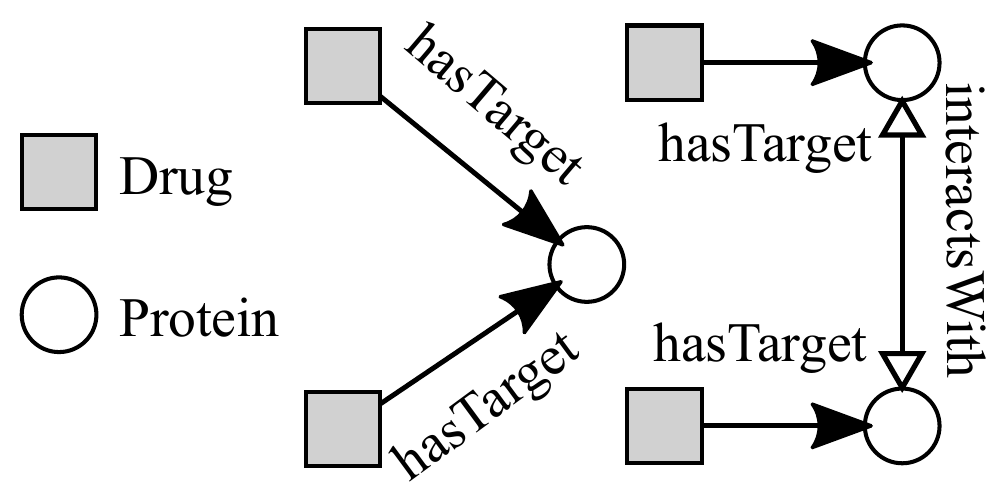}%
}{%
  \caption{\label{fig-relational-feature-types}Types of relational features.}%
}
\end{floatrow}
\vspace{-7mm}
\end{figure}

We use the publicly-available, preprocessed version of the dataset used in~\cite{Zitnik2018}.\footnote{Available at \|{http://snap.stanford.edu/decagon}} It consists of a multi-relational knowledge graph with two main components: a protein-protein
and a drug-drug interaction network. Known drug-protein target relationships
connect these different components. The protein-protein interactions are derived from several existing sources; it
is filtered to include only experimentally-validated physical interactions in
human. The drug-drug interactions are extracted from the \twosides database~\cite{Tatonetti2012}.
The drug-protein target relationships are experimentally-verified interactions
from the \stitch~\cite{Szklarczyk2016} database. Finally, the \sider~\cite{Kuhn2016} and
\offsides~\cite{Tatonetti2012} databases were used to identify \monosideeffects
of each drug. Please see Table~\ref{tab:graph-size} for detailed statistics of
the size and density of each part of the graph. For more details, please see~\cite{Zitnik2018}.
Each drug-drug link corresponds to a particular \polysideeffect. Our goal will
be to predict missing drug-drug links.

\vspace{-3mm}

\section{Methods}
\label{sec:methods}

\vspace{-3mm}

%\subsection{Knowledge Graph Embedding Methods}

KG embedding methods learn vector representations for entities and relation types of a KG~\cite{Bordes2013}. We investigate the performance of \textsc{DistMult}~\cite{Yang2015}, a commonly-used KG embedding method whose symmetry assumption is well-suited to this problem due to the
symmetric nature of the drug-drug (\polysideeffect) relation type. The advantage of KG embedding methods are their efficiency and their ability to learn fine-grained entity types suitable for downstream tasks without hand-crafted rules.% or mined logical rules. 
These embedding methods, however, are less interpretable than rule-based approaches and cannot incorporate domain knowledge. 
%\subsection{Relational Features for Polypharmacy}

A \emph{relational feature} is a logical rule which is evaluated in the KG to determine its truth value. For instance, the formula $(\mathtt{drug_1}, \mathtt{hasTarget}, \mathtt{protein_1}) \wedge (\mathtt{drug_2}, \mathtt{hasTarget}, \mathtt{protein_1})$ corresponds to a binary feature which has value $1$ if both $\mathtt{drug_1}$ and $\mathtt{drug_2}$ have $\mathtt{protein_1}$ as a target, and $0$ otherwise. 
In this work, we leverage relational features modeling drug targets with the relation type $\mathtt{hasTarget}$ and protein-protein interactions with the relation type $\mathtt{interactsWith}$. Figure~\ref{fig-relational-feature-types} depicts the two features types we use in our polypharmacy model. For a pair of entities $(\mathtt{h},\mathtt{t})$, the relational feature vector is denoted by $\mathtt{r}_{(\mathtt{h},\mathtt{t})}$.
Relational features capture concrete relationships between entities; thus, as shown in Section~\ref{sec:results}, they offer explanations for our predictions.

%\subsection{Combining Latent and Relational Features}

\kblrn is a recently proposed framework for end-to-end learning of knowledge graph representations~\cite{GarciaDuran2018}. It learns a product of experts (PoE)~\cite{hinton2002training}  where each expert is responsible for one feature type. In the context of KG representation learning, the goal is to train a PoE that assigns high probability to true triples and low probabilities to triples assumed to be false. Let $\mathtt{d} = (\mathtt{h}, \mathtt{r}, \mathtt{t})$ be a triple.
The specific experts we use are defined as \\
\vspace{-2mm}

\begin{tabular}{lr}
$ \hspace{-6mm} f_{(\mathtt{r},\mathtt{L})}(\mathtt{d} \mid \theta_{(\mathtt{r},\mathtt{L})}) = \left\{
              \begin{array}{ll}
                \exp(\left(\mathbf{e}_{\mathtt{h}} * \mathbf{e}_{\mathtt{t}}\right)\cdot \mathbf{w}^{\mathtt{r}})  \\
                1 \mbox { for all } \mathtt{r'} \neq \mathtt{r} 
              \end{array}
            \right.
$ & and $f_{(\mathtt{r},\mathtt{R})}(\mathtt{d} \mid \theta_{(\mathtt{r},\mathtt{R})}) = \left\{
              \begin{array}{ll}
                \exp\left(\mathbf{r}_{(\mathtt{h},\mathtt{t})} \cdot \mathbf{w}^{\mathtt{r}}_{\mathtt{rel}}\right)  \\
                1 \mbox { for all } \mathtt{r'} \neq \mathtt{r} 
              \end{array}
            \right.
$ 
\end{tabular} \\
\noindent
where $*$ is the element-wise product, $\cdot$ is the dot product, $\mathbf{e}_{\mathtt{h}}$ and $\mathbf{e}_{\mathtt{t}}$ are the embedding of the head and tail entity, respectively, and $\mathbf{w}^{\mathtt{r}}, \mathbf{w}^{\mathtt{r}}_{\mathtt{rel}}$ are the parameter vectors for the embedding and relational features for relation type $\mathtt{r}$. 
The probability of triple $\mathtt{d} = (\mathtt{h}, \mathtt{r}, \mathtt{t})$ is now
\vspace{-3mm}
$$p( \mathtt{d} \mid \boldsymbol{\theta}) = \frac{ f_{(\mathtt{r}, \mathtt{L})}(\mathtt{d} \mid \theta_{(\mathtt{r},\mathtt{L}) })\ f_{(\mathtt{r}, \mathtt{R})}(\mathtt{d} \mid \theta_{(\mathtt{r},\mathtt{R})})}{\sum_{\mathtt{c}} f_{(\mathtt{r}, \mathtt{L})}(\mathtt{c} \mid \theta_{(\mathtt{r},\mathtt{L}) })\ f_{(\mathtt{r}, \mathtt{R})}(\mathtt{c} \mid \theta_{(\mathtt{r},\mathtt{R})})},$$
\vspace{-4mm}

\noindent
where $\mathtt{c}$ indexes all possible triples.
%Product of experts are usually trained with contrastive divergence (CD)~\cite{hinton2002training}.
As proposed in previous work, we approximate the gradient of the log-likelihood by performing negative sampling~\cite{GarciaDuran2018}.

\section{Experimental results}
\label{sec:results}

We now empirically evaluate our proposed approach based on multi-relational
knowledge graph completion to predict \polysideeffects.

\vspace{-4mm}

\paragraph{Dataset construction}
We follow the common experimental design previously used~\cite{Zitnik2018} to
construct our dataset. The knowledge graph only contains ``positive'' examples
for which \polysideeffects exist. Thus, we create a set of negative
examples by randomly selecting a pair of drugs and a \polysideeffect which does
not exist in the knowledge graph. We ensure that the number of positive and
negative examples of each \polysideeffect are equal.
%Thus, each record in our
%final dataset includes a pair of nodes, a \polysideeffect, and a binary label
%of whether that combination occurs in the ground truth KG.
We
then use stratified sampling to split the records in training, validation and
testing sets.

We use an instance of the relational feature types 
depicted in Figure~\ref{fig-relational-feature-types} if it occurs at least $10$ times in the KG.
We choose these relational feature types because they offer a biological explanation for \polysideeffects;
namely, a \polysideeffect may manifest due to unexpected combinations or interactions on the drug targets.

\vspace{-4mm}

\paragraph{Baselines}
We first compare our proposed approach to 
\decagon~\cite{Zitnik2018}. Second, we consider each drug as a binary
vector of indicators for each \monosideeffect and gene target. We construct
training, validation and testing sets by concatenating the vectors of
the pairs of drugs described above.
%This is then a multi-label prediction
%problem in which
We predict the likelihood of each \polysideeffect given
the concatenated vectors.

\vspace{-4mm}

\paragraph{Complete \decagon dataset}
We first consider the same setting considered previously~\cite{Zitnik2018}.
As shown in Table~\ref{tab:performance}(top), our simple
baseline, \distmult, and \kblrn all outperform \decagon. 
%We suspect that the relational features cause \kblrn to overfit during training, which explains \distmult's somewhat better performance.

\vspace{-4mm}

\paragraph{Drug-drug interactions only}
Next, we evaluate \polysideeffect prediction based
solely on the pattern of other \polysideeffects. Specifically, we completely
remove the drug-protein targets and protein-protein interactions from the KG;
thus, we use only the drug-drug \polysideeffects in the training set
for learning. We focus on \distmult and \kblrn since they outperformed
the other methods in the first setting.

Surprisingly, the results in Table~\ref{tab:performance}(middle)
show that both \distmult and \kblrn perform roughly the same (or even
\emph{improve} slightly) in this setting, despite discarding
presumably-valuable drug target information. However, as shown in
Table~\ref{tab:graph-size}, few drugs have annotated protein
targets. 
Thus, we hypothesize that the learning algorithms ignore this information due to its sparsity. 

\vspace{-4mm}

\paragraph{Drugs with protein targets only}
To test this hypothesis, we remove all drugs which do not
have any annotated protein targets from the KG (and the
associated triples from the dataset). That is, the drug target
information is no longer ``sparse'', in that all drugs
in the resulting KG have protein targets.

The results in Table~\ref{tab:performance}(bottom)
paint a very different picture than before; \kblrn significantly
outperforms \distmult. These results show that the combination
of learned (or embedding) features and relational features can
significantly improve performance when the relational features
are present in the KG.

\vspace{-4mm}

\paragraph{Explanations and hypothesis generation}
The relational features allow us
to explain predictions and generate new hypotheses for wet lab
validation.
We chose one of our high-likelihood predictions and
``validated'' it via literature evidence. In particular, the
ranking of the drug combination
\|{CID115237} (paliperidone) and \|{CID271} (calcium) for the side
effect ``pain''
increased from $24~223$ when using only the embedding features
(of $58~029$ pairs of drugs for which ``pain''
is not a known side effect)
to a top-ranked pair when also using the relational features.
Inspection of the relational features shows that the interaction
between lysophosphatidic acid receptor 1 (LPAR1) and matrix
metallopeptidase 2 (MMP2) is particularly important for this
prediction. The MMP family is known to be associated with
inflammation (pain)~\cite{Manicone2008}. Independently, calcium
already upregulates
MMP2~\cite{Munshi2002}. Paliperidone upregulates LPAR1, which
in turn has been shown to promote MMP activiation~\cite{Fishman2001}.
Thus, palperidone indirectly exacerbates the up-regulation of
MMP2 already caused by calcium; this, then, leads to increased pain.
Hence, the literature confirms our prediction discovered
due to the relational features.

\vspace{-3mm}

\section{Discussion}

\vspace{-3mm}

We have shown that multi-relational knowledge graph completion
can achieve state-of-the-art performance on the \polysideeffect prediction
problem. Further, relational features offer explanations for our
predictions; they can then be validated via the literature
or wetlab. In the future, we plan to extend this work by considering
additional features of nodes in the graph, such as
Gene Ontology annotations
for the proteins and chemical structure of the drugs.

\vspace{-5mm}

\begin{table}[H]
\small
\centering
\caption{{\small The performance of each approach on the pre-defined test set. The measures are: area under the receiver operating characteristic curve (AuROC), area under the precision-recall curve (AuPR), and the average precision for the top $50$ predictions for each polypharmacy side effect (AP@50). The best result within each group is in bold.}} \label{tab:performance}
\begin{tabular}{l|rrr}
\textbf{Method}                        & \multicolumn{1}{c}{\textbf{AuROC}} & \multicolumn{1}{c}{\textbf{AuPR}} & \multicolumn{1}{c}{\textbf{AP@50}} \\ \hline
Baseline                                      & 0.896 & 0.859 & 0.812 \\
\decagon (values reported in~\cite{Zitnik2018})                    & 0.872 & 0.832 & 0.803 \\
\distmult                      & \textbf{0.923} & \textbf{0.898} & \textbf{0.899} \\
\kblrn                & 0.899 & 0.878 & 0.857 \\ \hline
\distmult (drug-drug interactions only)            & \textbf{0.931} & \textbf{0.909} & \textbf{0.919} \\
\kblrn (drug-drug interactions only)      & 0.894 & 0.886 & 0.892 \\ \hline
\distmult (drugs with protein targets only)       & 0.534 & 0.545 & 0.394 \\
\kblrn (drugs with protein targets only) & \textbf{0.829} & \textbf{0.797} & \textbf{0.774}
\end{tabular}
\end{table}

 \bibliographystyle{splncs04}
 \bibliography{library}
\end{document}

%% file: latex-includes.tex
\usepackage{xspace}

\usepackage[dvipsnames]{xcolor}

\usepackage{xparse}

\usepackage{multicol}
\usepackage{framed}

\renewcommand{\(}{\begin{columns}}
\renewcommand{\)}{\end{columns}}
\newcommand{\<}[1]{\begin{column}{#1}}
\renewcommand{\>}{\end{column}}

\usepackage{algorithm}
\usepackage{algorithmicx}
\usepackage[noend]{algpseudocode}
\usepackage{amsmath,amssymb,bm}
\usepackage{mathtools}

\usepackage{pifont}% http://ctan.org/pkg/pifont
\DeclareMathSymbol{\naf}{\mathord}{symbols}{"18}

\DeclarePairedDelimiterX{\infdivx}[2]{(}{)}{%
  #1\;\delimsize|\delimsize|\;#2%
}

\DeclareDocumentCommand \expectation { o m } {%
  \ensuremath{\mathbb{E}%
  \IfValueTF {#1} {%
    _{#1} \left[ #2 \right]%
  }{%
    \left[ #2 \right]%
  }%
  }\xspace%
}

\makeatletter
\newcommand*{\indep}{%
  \mathbin{%
    \mathpalette{\@indep}{}%
  }%
}
\newcommand*{\nindep}{%
  \mathbin{%                   % The final symbol is a binary math operator
    \mathpalette{\@indep}{/}%
  }%
}
\newcommand*{\@indep}[2]{%
  \sbox0{$#1\perp\m@th$}%        box 0 contains \perp symbol
  \sbox2{$#1=$}%                 box 2 for the height of =
  \sbox4{$#1\vcenter{}$}%        box 4 for the height of the math axis
  \rlap{\copy0}%                 first \perp
  \dimen@=\dimexpr\ht2-\ht4-.2pt\relax
  \kern\dimen@
  \ifx\\#2\\%
  \else
    \hbox to \wd2{\hss$#1#2\m@th$\hss}%
    \kern-\wd2 %
  \fi
  \kern\dimen@
  \copy0 %                       second \perp
}
\makeatother

\def\|#1{\ensuremath{\mathtt{#1}}}
\def\!#1{\ensuremath{\mathbf{#1}}}
\def\*#1{\ensuremath{\mathcal{#1}}}